\documentclass{article}

\usepackage{epsfig}
\usepackage{amsmath}
\usepackage{comment}
\usepackage{fullpage}
\usepackage{color}
\usepackage{enumerate}
\usepackage{amssymb}
\usepackage{mathpazo}
\usepackage{mathtools}
\usepackage[numbers,sort&compress,comma]{natbib}
\usepackage[bf,footnotesize]{caption}

\newcommand{\eq}[1]{Eq.~(\ref{eq:#1})}

\newcommand{\eql}[2]{Eq.~(\ref{eq:#1})\,-\,(\ref{eq:#2})}

\newcommand{\fig}[1]{Fig.~\ref{fig:#1}}

\newcommand{\bpanel}[1]{\textbf{\panel{#1}}}
\newcommand{\panel}[1]{\textsf{#1}}

\title{\bf{Frequency-dependent returns in nonlinear public goods games}}
\author{Christoph Hauert$^{1,2,}$\footnote{Correspondence should be addressed to Christoph Hauert (\texttt{hauert@math.ubc.ca}).}\,\ \& Alex McAvoy$^{3, 4}$ \\[1.5em]
{\small $^1$ Department of Mathematics, University of British Columbia, Vancouver B.C. Canada, V6T 1Z2}\\
{\small $^2$ Department of Zoology, University of British Columbia, Vancouver B.C. Canada, V6T 1Z4}\\
{\small $^3$ School of Data Science and Society, University of North Carolina at Chapel Hill, Chapel Hill, NC 27599 USA}\\
{\small $^4$ Department of Mathematics, University of North Carolina at Chapel Hill, Chapel Hill, NC 27599 USA}}
\date{}

\begin{document}

\maketitle

\begin{abstract}
When individuals interact in groups, the evolution of cooperation is traditionally modeled using the framework of public goods games. These models often assume that the return of the public good depends linearly on the fraction of contributors. In contrast, in real life public goods interactions, the return can depend on the size of the investor pool as well. Here, we consider a model in which the multiplication factor (marginal per capita return) for the public good depends linearly on how many contribute, which results in a nonlinear model of public goods. This simple model breaks the curse of dominant defection found in linear public goods interactions and gives rise to richer dynamical outcomes in evolutionary settings. We provide an in-depth analysis of the more varied decisions by the classical rational player in nonlinear public goods interactions as well as a mechanistic, microscopic derivation of the evolutionary outcomes for the stochastic dynamics in finite populations and in the deterministic limit of infinite populations. This kind of nonlinearity provides a natural way to model public goods with diminishing returns as well as economies of scale.
\end{abstract}

\section{Introduction}
Cooperation under Darwinian selection is a recurring theme in evolutionary theory. The prisoner's dilemma is a leading mathematical metaphor to study the evolution of cooperation. In a particularly simple and intuitive instance of the prisoner's dilemma known as the donation game \citep{sigmund:PUP:2010}, a cooperator provides a benefit, $b$, to a co-player at a cost, $c$, to themselves, where $b>c$. Thus, if both individuals cooperate they each get $b-c$ but if both evade the costs of cooperation and defect, neither gets anything. Since $b-c>0$ they both prefer mutual cooperation over mutual defection. However, a defector interacting with a cooperator shirks the costs but reaps the benefit $b$, while the cooperator is saddled with the cost, $c$. Hence, it pays to defect regardless of the opponent's strategy. In fact, the change in payoff is independent of the opponent's strategy, a property termed ``equal-gains-from-switching'' \citep{nowak:AAM:1990}. Thus, defection is a dominant strategy and, consequently, mutual defection the sole Nash equilibrium. As a result, a conflict of interest ensues between the individuals and the pair, which is the hallmark of social dilemmas \citep{dawes:ARP:1980,hauert:JTB:2006a}. Individual interests undermine cooperation to the detriment of all.

In the closely-related snowdrift game \citep{sugden:Blackwell:1986,hauert:Nature:2004,doebeli:EL:2005} cooperators get a share of the benefits. In a common formulation, both players get a benefit, $b$, as long as at least one of them is a cooperator. If exactly one player is a cooperator, then this individual incurs a cost of $c$. If both cooperate, then the two players share this cost, each bearing $c/2$. In other words, a cooperator gets $b-c/2$ against a cooperator and $b-c$ against a defector; and a defector gets $b$ against a cooperator and $0$ against a defector. As long as $b<c<2b$ the characteristics of the interaction are unchanged and recover the prisoner's dilemma. For sufficiently large benefits, $b>c$, however, the best strategy depends on the opponent's decision. Defection is still the better option against a cooperator but against a defector it now pays to switch to cooperation. Thus, the social dilemma is relaxed: even though the social optimum of mutual cooperation remains prone to cheating, defection is no longer dominant and cooperation no longer invariably doomed.

These games are, of course, typically understood as two-player interactions. In the following, we consider public goods games with non-constant multiplication factors. We demonstrate that simple nonlinearities arising in this way recover the interaction characteristics of the prisoner's dilemma and the snowdrift game (and more), for more general kinds of interactions in groups of size $n$. We first discuss the choices of rational, payoff-maximizing players in both linear and nonlinear settings. In subsequent sections, we turn to evolutionary dynamics in the deterministic limit of infinite populations, as well as stochastic scenarios in finite populations with demographic fluctuations.

\subsection{Linear public goods}
Public goods games represent a generalized model for studying the evolution of cooperation in larger groups: cooperators contribute a fixed amount, $c$, to a common pool, while defectors contribute nothing. The total contributions are multiplied by a factor $r$, representing the marginal per capita return, and equally divided among all participants, regardless of whether or not they contributed. Thus, the respective payoffs for defectors, $\pi_D$, and cooperators, $\pi_C$, in a group of size $n$ with $k$ cooperators (and $n-k$ defectors) are
\begin{subequations}
	\label{eq:linpgg}
	\begin{align}
		\pi_D\left(k\right) &= \frac{kr}n c;\\
		\pi_C\left(k\right) &= \pi_D\left(k
		\right) - c =\left(\frac{kr}n-1\right)c.
	\end{align}
\end{subequations}
(Note that $\pi_{D}\left(k\right)$ makes sense only when $k<n$; similarly, $\pi_{C}\left(k\right)$ is defined when $k>0$.) Universal cooperation pays $\pi_C\left(n\right)= \left(r-1\right) c$, while universal defection pays nothing, $\pi_D\left(0\right) =0$. For $r>1$, everyone prefers cooperation to defection. Nevertheless, defectors always outperform cooperators because they shirk the costly contributions to the common pool, $\pi_D\left(k\right) >\pi_C\left(k\right)$ for any $k\in\left\{1,\dots ,n-1\right\}$. Hence, defection remains the dominant strategy just as it is in the prisoner's dilemma. Maximizing individual payoffs happens to the detriment of everyone else. As a consequence, if everyone maximizes their (short-term) gains, contributions to the common pool dwindle and participants forgo the benefits of the public good, again recovering a social dilemma.

In the public goods game, \eq{linpgg}, the return of the common pool linearly increases with the fraction of contributors $k/n$ and, as a consequence, this scenario is often referred to as the \emph{linear} public goods game. The setup is closely related to the prisoner's dilemma, or, more precisely, the donation game. In fact, one public goods interaction in a group of size $n$ is equivalent to each participant engaging in $n-1$ pairwise interactions with co-players \citep{hauert:Complexity:2003}, at least when the population is unstructured. (In structured populations, it might be impossible for all members of a group to interact pairwise \citep{mcavoy:JMB:2016}, which can complicate the notion of reducibility of a game.) Games exhibiting such simplifying symmetries are termed additive games, which represents a natural generalization of the ``equal-gains-from-switching'' property in pairwise interactions.

\subsection{Nonlinear public goods}
In nature, linear public goods games are likely the exception and more often the return of the public good will depend on the total contributions to the common pool. As \citet{archetti:JTB:2011} note, ``no linear public goods have been reported, apart from artificial public goods in experimental settings for behavioral experiments.'' This applies regardless of whether the interaction refers to microscopic scales such as extracellular products in microbial organisms (e.g., hydrolyzing sucrose in yeast \citep{travisano:TREE:2004}, antibiotic resistance in bacteria \citep{neu:Science:1992}, and those involving communication mechanisms like quorum sensing \citep{brown:PRSB:2001,diggle:Nature:2007,czaran:PLOSONE:2009}), or macroscopic scales such as cooperative hunting \citep{stander:BES:1992} and conservation issues (e.g., fisheries management \citep{kraak:FF:2011,squires:MP:2014} and climate change \citep{seo:Elsevier:2017}). In particular, benefits of larger pools may be synergistically enhanced (economies of scale) or discounted due to inefficiencies or effects of saturation (diminishing returns) \citep{hauert:JTB:2006a}.

The natural prevalence of nonlinear public goods stands in stark contrast to the overwhelming focus on linear public goods in evolutionary theory as well as behavioural experiments. Nonlinear public goods games have received comparatively less attention \citep{archetti:Evolution:2011}. Notable examples in the literature include threshold public goods \citep{bagnoli:EI:1991,cadsby:JPE:1999,wang:CSF:2024} and collective risk dilemmas \citep{milinski:PNAS:2008,santos:PNAS:2011}, which require minimum investment levels for the public good to be realized. In such cases, the multiplication factor corresponds to a step-function, which is zero if contributions fall below a threshold and constant if the threshold is reached. In cancer cells, nonlinear (but non-threshold) public goods arise from sigmoidal functions of concentration of growth factors \citep{archetti:EA:2013,archetti:JTB:2016} such as IGF-II in neuroendocrine pancreatic cancer \citep{archetti:PNAS:2015}.

Here, we consider a model of nonlinear public goods based on a multiplication factor, $r\left(k\right)$, which depends on the number of contributors, $k$. The function $r\left(k\right)$ can (in principle) be any mapping from integers $k\in\left\{0,\ldots,n\right\}$ to real numbers, which includes threshold returns such as those encountered in crowdfunding \citep{agrawal:JEMS:2015}. However, we focus on a particular form of $r\left(k\right)$, which yields a simple, one-parameter model of nonlinearity in public goods. For increasing functions of $k$, this models economies of scale \citep{bejan:JAP:2017}, while decreasing functions describe diminishing returns \citep{brue:JEP:1993}, and constant $r\left(k\right)\equiv r$ recovers the traditional, linear case. The payoffs for defectors and cooperators become
\begin{subequations}
	\begin{align}
		\pi_D\left(k\right) &= \frac{kr\left(k\right)}n c ;\\
		\pi_C\left(k\right) &= \pi_D\left(k\right)- c =\left(\frac{kr\left(k\right)}n -1\right)c.
	\end{align}
\end{subequations}
As before, defectors are invariably better off than cooperators since
\begin{align}
	\label{eq:pckpdk}
	\pi_C\left(k\right)-\pi_D\left(k\right) = -c < 0
\end{align}
for any $k\in\left\{1,\dots ,n-1\right\}$, independent of the choice of $r\left(k\right)$. However, introducing nonlinearities can change the characteristics of the interaction such that universal defection is no longer the inevitable outcome. More specifically, if the increase in return of the common pool arising from one additional contributor exceeds the costs of contributing, meaning
\begin{align}
	\label{eq:pck1pdk}
	\pi_C\left(k+1\right)-\pi_D\left(k\right) > 0, \quad\text{or, equivalently, }\quad \left(k+1\right) r\left(k+1\right)-k r\left(k\right) > n,
\end{align}
then it indeed pays to switch to cooperation. The condition \eq{pck1pdk} relaxes the social dilemma because rational individuals switch to cooperation despite the fact that the other members of the group profit even more than the individual itself. The flip side is that switching to defection becomes an act of spite because an individual accepts a reduction in its payoff only to harm everyone else to a greater extent \citep{smead:Evolution:2013}. This represents a generalization of the snowdrift game for pairwise interactions to interactions in groups.

\subsection{Linear multiplication factors}
An interesting and illustrative example of a nonlinear public good arises if the multiplication factor, $r\left(k\right)$, linearly depends on the number of contributors, $k$. If the multiplication factor is $r_1$ for a single contributor in the group and $r_n$ in groups of only contributors, then
\begin{align}
	\label{eq:rkr1rn}
	r\left(k\right) = r_1 + \frac{k-1}{n-1} \left(r_n-r_1\right) .
\end{align}
Thus, the return of the common pool is quadratic in the number of contributors. Diminishing returns are characterized by $r_1>r_n$, while $r_1<r_n$ refers to economies of scale. 

Universal defection is not rational if there is an incentive to switch to cooperation in this state, meaning $\pi_C\left(1\right) >\pi_D\left(0\right)$ (or $r_1>n$). On the other hand, universal cooperation is rational if no agent has an incentive to defect in this state, meaning $\pi_C\left(n\right) >\pi_D\left(n-1\right)$ (or $2r_n-r_1>n$). It is possible that both universal cooperation and universal defection constitute rational outcomes -- or, conversely, that neither does. They are both rational if $2r_1<n+r_1<2r_n$, which requires $r_n>r_1$ and hence only applies in the case of economies of scale. In addition, this implies that a threshold, $k^\ast$, must exist such that for $k>k^\ast$ it pays to switch to cooperation, whereas for $k<k^\ast$ it pays to switch to defection. This switch happens when $\pi_C\left(k^\ast+1\right) =\pi_D\left(k^\ast\right)$, which yields a threshold of
\begin{align}
	\label{eq:kast}
	k^\ast = \left(n-1\right)\frac{n-r_1}{2\left(r_n-r_1\right)}.
\end{align}
Switching strategies increases (decreases) $k$. This results in a coordination game where it is always best to do what the majority does, only that what counts as a ``majority'' is determined by $k^\ast$. In pairwise interactions, this corresponds to a stag-hunt game \citep{skyrms:CUP:2003}.

Similarly, neither universal cooperation nor defection are rational if the converse holds ($2r_1>n+r_1>2r_n$), which implies $r_n<r_1$ and hence only applies in the case of diminishing returns. Again, a threshold $k^\ast$ must exist and is indeed again given by \eq{kast}. However, now it pays to switch to defection if $k>k^\ast$ and to cooperation if $k<k^\ast$. Such a switch decreases (increases) $k$. This results in a coexistence game where the rare type is favored, only that what counts as ``rare'' is given by $k^\ast$. In pairwise interactions, this corresponds to a snowdrift game (or, equivalently, a hawk-dove or chicken game if the interpretation of the interaction is competitive rather than cooperative).

Note that switching to cooperation can be rational and is not in violation of \eq{pckpdk} (which states that defectors are always better off than cooperators) because the switch changes the number of cooperators in the group and hence the return of the common pool. In fact, cooperation can even be a dominant strategy if $\pi_C\left(k+1\right)>\pi_D\left(k\right)$ holds for all $k$. For linear $r\left(k\right)$ it is enough to ensure $\pi_C\left(1\right)>\pi_D\left(0\right)$ as well as $\pi_C\left(n\right) >\pi_D\left(n-1\right)$, which translates to $r_1>n$ and $2r_n-r_1>n$ and simplifies to $r_1,r_n>n$. Conversely, defection is dominant with all inequalities reversed. An alternative but equivalent dominance condition is $k^\ast\notin\left[0,n\right]$, although further checks are required to determine which strategy is dominant.

A convenient parametrization is to set $r_1 = r-a$ and $r_n = r+a$, where $r$ denotes the multiplication factor in the traditional, linear case and $a$ reflects the nonlinearity. For $a<0$ the public good provides diminishing returns, whereas $a>0$ represent economies of scale. The contributor threshold $k^\ast$ then becomes
\begin{align}
	k^\ast &= \left(n-1\right)\frac{n-r+a}{4a}.
\end{align}
Interestingly, in the special case of cost-free contributions, $r=n$, this reduces to $k^\ast =\left(n-1\right) /4$. Thus, if at least one quarter of the $n-1$ co-players are cooperating, then it pays to switch to cooperation for synergistic public goods (economies of scale) and at most one quarter for discounted public goods (diminishing returns). This observation complements the $1/2$-rule of risk dominance \citep{harsanyi:GEB:1995} and the $1/3$-rule of evolutionary dynamics \citep{nowak:Nature:2004,ohtsuki:JTB:2007c} for pairwise interactions.

\section{Evolutionary dynamics}
Evolutionary game theory considers populations of individuals that are characterized by their traits (strategies). All individuals accrue payoffs from interactions with other members of the population. Following the principles of Darwinian selection, individuals with high payoffs have either \emph{(i)} higher chances to produce offspring, which inherit the parental strategy, or \emph{(ii)} higher propensities that their strategy is imitated by others. As a consequence, more successful strategies tend to spread in the population.

Complex features of individuals, such as fecundity, or some other measure of success (``fitness'') almost certainly depend on numerous factors. The limit of weak selection assumes that the impact of one particular component under consideration reflects a small perturbation of an averaged trait. Thus, weak selection is not only biologically and socially meaningful but also mathematically convenient because it renders analytical considerations more tractable and intuitively more accessible.

\subsection{Payoffs and fecundity}
In general, payoffs can be any real number but are usually translated into fecundity or other non-negative measures of fitness. The most versatile mapping is an exponential map for strategic type $i\in\left\{C,D\right\}$, with expected payoff $u_i$ mapped to $f_i = e^{wu_i}$, where $w\geq 0$ denotes the selection strength. This mapping is particularly convenient because \emph{(i)} it does not limit the strength of selection, \emph{(ii)} it ensures $f_i>0$, which simplifies conversions to probabilities, and \emph{(iii)} in the limit of weak selection recovers another popular mapping $f_i=1+wu_i$, where $1$ denotes a static (and normalized) baseline fecundity or fitness. Even under stronger selective pressures, this mapping is natural when competition does not depend on the shared baseline payoff in a game \citep{mcavoy:PLOSCB:2021}.

Note that the expected payoffs, $u_i$, also depend on the procedure for sampling interaction groups, which may, for example, depend on the population structure \citep{hauert:PRE:2018}. In the simplest case, the population is unstructured (or well-mixed), which means that every individual is equally likely to interact with every other member of the population. As a consequence, the payoff to each individual only depends on the abundance of the different strategic types in the population, or, in the limit of infinite populations, on their frequencies.

Moreover, depending on the parameters and population configurations, situations can occur where the expected payoff to cooperators exceeds that of defectors, $u_C>u_D$. Nevertheless, this does not contradict the fact that defectors consistently outperform cooperators, $\pi_C\left(k\right) <\pi_D\left(k\right)$, for any mixed group, $k=1,\ldots,n-1$. Instead, this finding represents an instance of Simpson's paradox \citep{wagner:AmStat:1982} in the sense that even though defectors are better off in mixed groups, on average cooperators can offset their losses through interactions in groups of only cooperators ($k=n$) -- and, similarly, on average defectors are disadvantaged by interactions in groups of only defectors ($k=0$).

\subsection{Strategy dynamics}
Evolution through asexual reproduction is conveniently modeled through the repeated processes of births and deaths \citep{doebeli:eLife:2017}. The case where population size remains constant, death occurs uniformly at random (i.e., all individuals have the same life expectancy), and births are proportional to fecundity or fitness, is captured by the (frequency dependent) Moran process \citep{moran:PCPS:1958,nowak:Nature:2004}. Conversely, in cultural evolution, a focal individual probabilistically compares its payoff to that of other members in the population. This is termed a ``pairwise comparison'' process when considering only a single model member \citep{traulsen:PRE:2012}. Naturally, there are numerous ways to implement this probabilistic comparison. However, a particularly convenient form for the probability that a focal individual of type $i$ and fitness $f_i$ adopts the type $j$ of a model with fitness $f_j$ is given by
\begin{align}
	\label{eq:paircomp}
	\omega\left(f_i,f_j\right) = \frac12\left(1+\frac{f_j-f_i}{f_i+f_j}\right), \quad\text{or, equivalently, }\quad \omega\left(f_i, f_j\right) = \frac{f_j}{f_i+f_j}.
\end{align}
This means the focal individual adopts the model's type $j$ according to a biased coin toss. The positive or negative bias is given by their normalized difference in performance. For the exponential payoff-to-fitness mapping, this formulation recovers the popular Fermi update rule,
\begin{align}
	\label{eq:fermi}
	\omega\left(u_i,u_j\right) = \frac{1}{1+e^{-w \left(u_j-u_i\right)}} ,
\end{align}
where $u_i$ and $u_j$ refer to the expected payoffs of individuals of type $i$ and $j$, respectively. The limit of weak selection, $w\ll1$, results in the linear comparison
\begin{align}
	\label{eq:fermiweak}
	\omega\left(u_i, u_j\right) = \frac{1}{2} \left(1+\frac{w}{2} \left(u_i-u_j\right)\right) +O\left(w^{2}\right) .
\end{align}
In all cases, the dynamics are determined by the probabilities that the number of strategies increases or decreases by one. More specifically, for the frequency dependent Moran process, the probabilities $T_i^+$ to increase the number of cooperators $X\to X+1$, or $T_i^-$ to decrease them, $X\to X-1$, in a population with $i$ cooperators are given by
\begin{subequations}
	\label{eq:t+t-moran}
	\begin{align}
		T_i^+ &= \frac{if_C}{if_C + \left(N-i\right) f_D}\frac{N-i}{N} ;\\
		T_i^- &= \frac{\left(N-i\right) f_D}{if_C + \left(N-i\right) f_D}\frac{i}{N},
	\end{align}
\end{subequations}
where $f_C$ and $f_D$ denote the expected fitness values of cooperators and defectors, respectively. The first term in \eq{t+t-moran} indicates the probability that a cooperator (defector) produces a clonal offspring, and the second term denotes the probability that the offspring replaces a defector (cooperator). Similarly, for the pairwise comparison process, the transition probabilities are 
\begin{subequations}
	\label{eq:t+t-pair}
	\begin{align}
		T_i^+ &= \frac{i}{N}\frac{N-i}{N}\omega\left(u_D,u_C\right) ; \\
		T_i^- &= \frac{N-i}{N}\frac{i}{N}\omega\left(u_C,u_D\right) .
	\end{align}
\end{subequations}
The first two terms indicate the probabilities that the focal individual is of one type and the model of the other. All other combinations do not result in a change of the population composition. The last term then denotes the probability that the focal individual adopts the model's strategy.

\subsection{Mutation and errors}
So far, we have assumed that no mutants arise and no mistakes happen. However, mutational processes or noisy imitation can be easily accommodated in modified transition probabilities. In either case, a microscopic description of the process is required. For asexually reproducing populations, mutants are typically introduced through either ``cosmic rays'' such that any individual may spontaneously adopt another strategy (uniform mutations) or through random errors during reproduction (fitness or temperature-based mutations) \citep{mcavoy:PLOSCB:2021}. Similarly, for cultural evolution (pairwise comparison) the focal individual can either spontaneously switch to another strategy (random exploration) \citep{traulsen:PNAS:2009} or make mistakes when assessing or adopting the model's strategy (perception and implementation errors). In all instances, the resulting, modified transition probabilities, $Q_i^\pm$, can be easily derived \citep{traulsen:PRE:2006a}. The simplest case is given by uniform mutations:
\begin{subequations}
	\label{eq:q+q-unif}
	\begin{align}
		Q_i^+ &= \left(1-\mu\right) T_i^+ + \mu\frac{N-i}N ;\\
		Q_i^- &= \left(1-\mu\right) T_i^- + \mu\frac{i}N,
	\end{align}
\end{subequations}
which states that with probability $1-\mu$ everything proceeds as before but with probability $\mu$ an individual randomly switches to the opposite strategy. Similarly, when mutants arise through errors during reproduction (also termed temperature-based mutations because ``hot'' individuals with higher fitness produce more mutants), the transition probabilities become
\begin{subequations}
	\label{eq:q+q-temp}
	\begin{align}
		Q_i^+ &= (1-\mu)T_i^+ + \mu\frac{N-i}i T_i^- ;\\
		Q_i^- &= (1-\mu)T_i^- + \mu\frac{i}{N-i} T_i^+ ,
	\end{align}
\end{subequations}
for $i\in\left\{1,\ldots,N-1\right\}$, together with $Q_0^+=\mu f_D/\left(f_C+f_D\right)$ and $Q_N^-=\mu f_C/\left(f_C+f_D\right)$, because homogeneous cooperator states produce more mutants due to their higher rates of reproduction than homogeneous defector states. The notable difference to \eq{q+q-unif} is that now type $i$ increases not only if an $i$ type reproduces and replaces one of the $N-i$ other types, but also if one of the $N-i$ other types reproduces, creates a mutant offspring and the mutant succeeds in replacing a parental type. Also note that $T_i^\pm$ already includes the probability that an individual had been replaced and hence needs to be multiplied by $N/\left(N-i\right)$ or $N/i$, respectively.

\section{Infinite populations}
In the limit of infinite populations, $N\to\infty$, the microscopic update rules recover the deterministic replicator equation (or variants thereof) \citep{hofbauer:CUP:1998,taylor:MB:1978,maynard-smith:CUP:1982} based on the transition probabilities $T_i^\pm$ or $Q_i^\pm$, \eq{t+t-moran}-\eq{q+q-temp} \citep{traulsen:PRL:2005}. More specifically, the dynamical equations follow from a system size expansion for the transition probabilities, such as \eq{t+t-moran} or \eq{t+t-pair}:
\begin{align}
	\label{eq:rep:inf}
	\dot x &= T^+\left(x\right) -T^-\left(x\right) ,
\end{align}
where $T^\pm\left(x\right)$ are the continuous analogues of $T_i^\pm$ with $i/N\to x$. In particular, for the Fermi update, \eq{fermi}, this limit yields
\begin{align}
	\label{eq:rep:fermi}
	\dot x &= x \left(1-x\right) \tanh\left(\frac{w}{2} \left(u_C - u_D\right)\right),
\end{align}
and recovers the replicator equation \citep{hofbauer:CUP:1998},
\begin{align}
	\label{eq:rep:fermiweak}
	\dot x &= x \left(1-x\right) \frac{w}{2} \left(u_C - u_D\right),
\end{align}
for weak selection (except for a constant rescaling of time, denoted by the factor $w/2$). For the Moran process, this limit recovers the adjusted replicator equation \citep{maynard-smith:CUP:1982},
\begin{align}
	\label{eq:rep:moran}
	\dot x &= x \left(1-x\right) \frac1{\cal F}\left(f_C-f_D\right),
\end{align}
where ${\cal F} = xf_C+\left(1-x\right) f_D$ denotes the average population fitness. The key difference from the standard replicator equation is that evolutionary changes happen fast if the population is performing poorly on average (${\cal F}$ small), but slow in populations that are well off (${\cal F}$ large). However, all equilibria and their stability remain unchanged. In the weak selection limit, fitness differences are small and hence ${\cal F}$ constant to first order. As a consequence, this limit again recovers the standard replicator equation (although changes happen twice as fast when compared to \eq{rep:fermiweak}).

Any of the above forms of the replicator equation admit up to three equilibria: the two trivial equilibria at $x^\ast_0=0$ and $x^\ast_1=1$, denoting the homogeneous states with all defectors and all cooperators, respectively, as well as possibly a third, interior equilibrium $x^\ast$ as the solution to $u_C=u_D$, provided $x^\ast\in\left(0,1\right)$. The existence and location of $x^\ast$ as well as the stability of all equilibria depend on the interaction parameters.

Including mutations through \eq{q+q-unif} or \eq{q+q-temp} results in various forms of replicator-mutator equations. The traditional form is recovered for uniform mutations in the limit of weak selection \citep{page:JTB:2002,traulsen:PRE:2006c},
\begin{align}
	\label{eq:repmut:inf}
	\dot x &= \left(1-\mu\right) x \left(1-x\right) \left(u_C - u_D\right) + \mu \left(1-2x\right) ,
\end{align}
with the selection strength absorbed in a constant rescaling of time.

\subsection{Linear multiplication factors}
For linearly increasing or decreasing multiplication factors, \eq{rkr1rn}, the average payoffs for defectors and cooperators in a well-mixed population with cooperator frequency $x$ (and $1-x$ defectors) and random (binomial) sampling of interaction groups are given by
\begin{subequations}
	\label{eq:sumCD}
	\begin{align}
		u_D &= \sum_{k=0}^{n-1} x^k \left(1-x\right)^{n-1-k} \binom{n-1}{k} \pi_D\left(k\right) \nonumber \\
		&= \frac{x c}n \left(\left(n-1\right) r_1 + \left(n-2\right) x \left(r_1-r_n\right)\right) ; \\
		u_C &= \sum_{k=0}^{n-1} x^k \left(1-x\right)^{n-1-k} \binom{n-1}{k} \pi_C\left(k+1\right) \nonumber \\
		&= u_{D} -\left(1-\frac{r_1+2 x \left(r_1-r_n\right)}n\right)c .
	\end{align}
\end{subequations}
The replicator dynamics are then described by
\begin{align}
	\dot x &= x\left(1-x\right)\left(\frac{r_1+2x\left(r_1-r_n\right)}n-1\right)c .
\end{align}
This equation has trivial equilibria at $x^\ast_0=0$ and $x^\ast_1=1$, as well as an interior equilibrium at
\begin{align}
	\label{eq:xast}
	x^\ast &= \frac{n-r_1}{2\left(r_n-r_1\right)},
\end{align}
provided that it exists (meaning $x^\ast\in\left(0,1\right)$).

Defection is stable if $r_1<n$ and cooperation is stable if $2r_n>n+r_1$. Thus, the interior equilibrium $x^\ast$ exists and is stable if $r_1>n>2r_n-r_1$, which requires diminishing returns ($r_n<r_1$). It is unstable if the inequalities are reversed, which requires economies of scale ($r_n>r_1$). Otherwise, no interior equilibrium exists and one of the strategies must be dominant. Note, incidentally, that $x^\ast$ relates to the threshold number of cooperators in the group, $k^\ast =\left(n-1\right) x^\ast$, above which rational players switch to cooperation (defection) if $x^\ast$ is unstable (stable).

\begin{figure}[tp]
	\centerline{\includegraphics[width=0.9\textwidth]{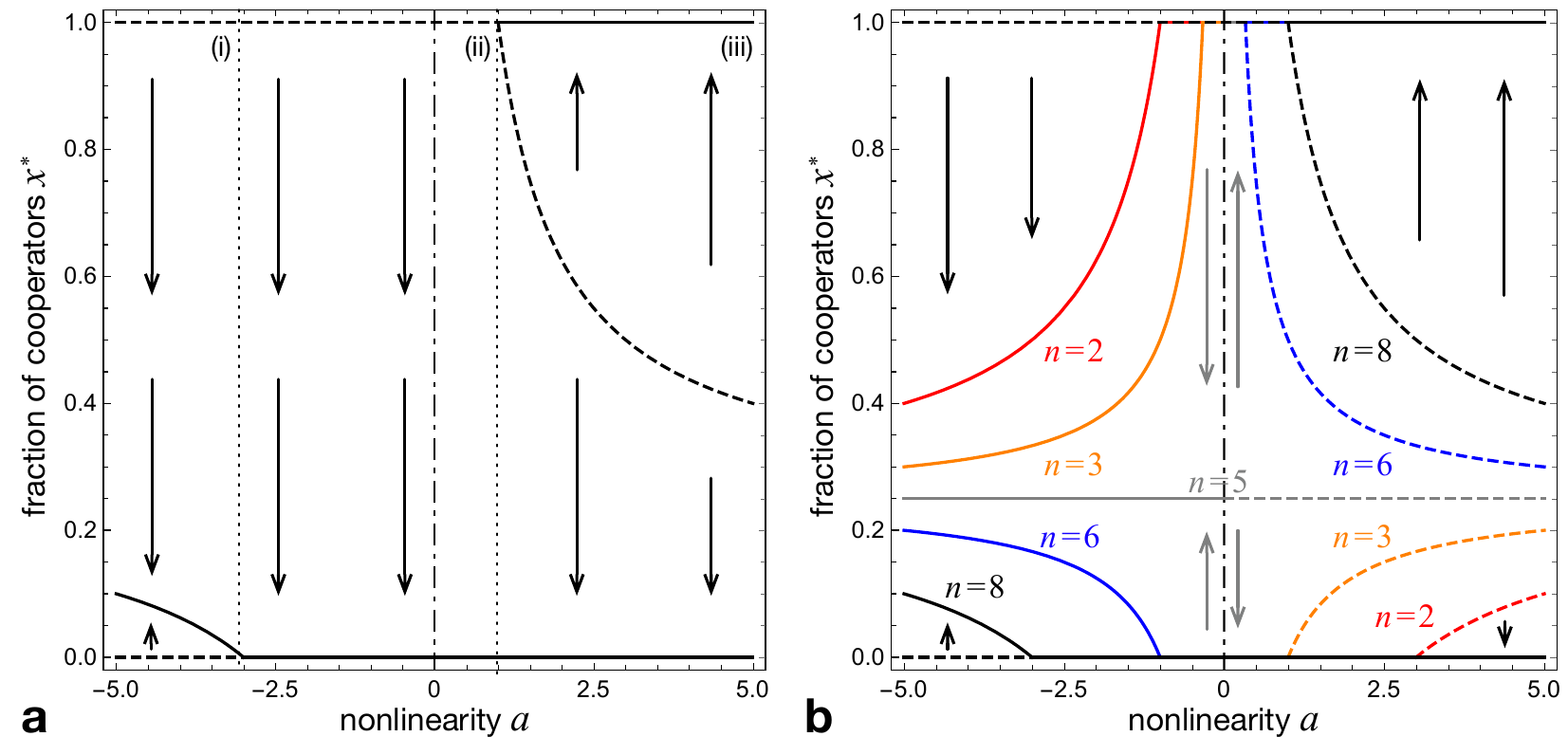}}
	\caption{\label{fig:dyn:cnpgg}
		Evolutionary dynamics in compulsory, nonlinear public goods interactions with $r_1=r-a$ and $r_n=r+a$, where $a$ denotes the strength of the nonlinearity (with $a>0$ representing economies of scale and $a<0$ representing diminishing returns, separated by the dash-dotted line).
		\textbf{\textsf{a}} Stable (solid lines) and unstable equilibria (dashed lines) are shown as a function of the nonlinearity, $a$, for $n=8$, $r=5$, and $c=1$. This results in three dynamical regimes (separated by dotted lines): 
		\emph{(i)} stable coexistence, 
		\emph{(ii)} dominance of defection, and 
		\emph{(iii)} bistability between homogeneous states where everyone ends up cooperating or defecting depending on the initial configuration.
		\textbf{\textsf{b}} is the same as \textbf{\textsf{a}} but for different interaction group sizes, $n$ (and again with $r=5$ and $c=1$). For $a<0$, the equilibrium frequency of cooperators decreases with $n$, while for $a>0$ the basin of attraction of cooperation decreases with $n$. For the special case of cost-free cooperation, $r=n$, the interior equilibrium $x^\ast =1/4$ is independent of the nonlinearity, $a$.
	}
\end{figure}

Using the parametrization $r_1 = r-a$ and $r_n = r+a$, \fig{dyn:cnpgg} shows that the nonlinearity $a$ gives rise to three dynamical regimes.
For heavily-discounted public goods ($a<r-n$), cooperators and defectors coexist at an equilibrium frequency $x^\ast =\left(n-r+a\right) /\left(4a\right)$, while for strongly-synergistic public goods ($a>\left(n-r\right) /3$), the dynamics are bistable with the respective basins of attractions separated by $x^\ast$. In between these two extremes ($r-n<a<\left(n-r\right)/3$ for $r<n$ and $\left(n-r\right) /3<a<r-n$ for $r>n$), no interior equilibrium $x^\ast$ exists. In this case, defection dominates for $r<n$ and cooperation dominates for $r>n$ (see \fig{dyn:cnpgg}\panel{b}). Finally, for cost-free cooperation ($r=n$), the interior equilibrium is fixed at $x^\ast =1/4$ and is stable for $a<0$ and unstable for $a>0$.

\begin{figure}[tp]
	\centerline{\includegraphics[width=0.9\textwidth]{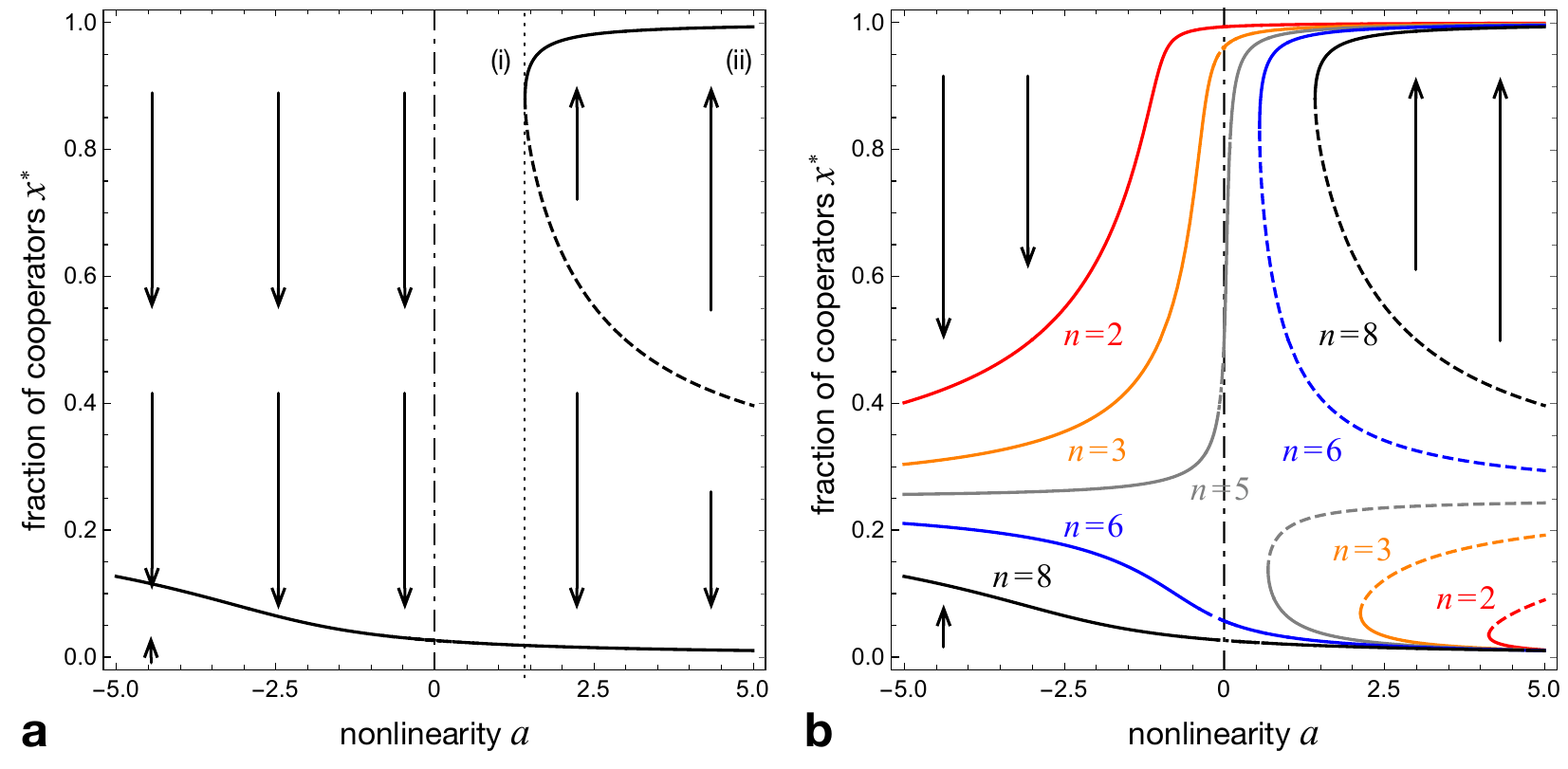}}
	\caption{\label{fig:dyn:cmnpgg}
		Evolutionary dynamics in nonlinear public goods interactions with uniform mutations. Here, $r_1=r-a$ and $r_n=r+a$, where $a$ denotes the strength of the nonlinearity (with $a>0$ giving economies of scale and $a<0$ giving diminishing returns, separated by the dash-dotted line).
		\bpanel{a} Stable (solid lines) and unstable equilibria (dashed lines) are shown as a function of the nonlinearity, $a$, for $n=8$, $r=5$, $c=1$, and $\mu=0.01$. Two dynamical regimes are observed (separated by the dotted line):
		\emph{(i)} stable coexistence for $a<0$ and extending into $a>0$ until 
		\emph{(ii)} a saddle node bifurcation gives rise to bistability through two additional interior equilibria.
		\bpanel{b} is the same as \bpanel{a} but for different interaction group sizes, $n$ (again using $r=5$, $c=1$, and $\mu=0.01$). In the region of coexistence, the equilibrium frequency of cooperators decreases with $n$, while for bistability the basin of attraction of cooperation decreases with $n$.
	}
\end{figure}

When including mutations, the trivial equilibria $x^\ast_0=0$ and $x^\ast_1=1$ cease to exist (see \fig{dyn:cmnpgg}).
Instead, for diminishing returns ($a<0$), only a single mixed equilibrium exists and, depending on the group size, $n$, may extend more or less into the realm of economies of scale, $a>0$. For larger $a$, a saddle node bifurcation introduces two additional interior equilibria that give rise to bistable dynamics (see \fig{dyn:cmnpgg}\panel{b}).

\section{Finite populations}
In a finite population of fixed size, $N$, the state of the population is determined by the number of cooperators, $X$ (with $N-X$ defectors). Evolutionary trajectories are then determined by a stochastic process that balances demographic noise arising from finite population sizes and selection based on fitness differences between cooperators and defectors. If interaction groups are sampled uniformly at random, the group composition follows a hypergeometric distribution. The expected payoffs of cooperators and defectors are then given by
\begin{subequations}
	\label{eq:randomsampling}
	\begin{align}
		u_C &= \sum_{k=0}^{n-1} \frac{\binom{X-1}{k} \binom{N-X}{n-1-k}}{\binom{N-1}{n-1}}\pi_C\left(k+1\right) ; \\
		u_D &= \sum_{k=0}^{n-1} \frac{\binom{X}{k} \binom{N-X-1}{n-1-k}}{\binom{N-1}{n-1}}\pi_D\left(k\right),
	\end{align}
\end{subequations}
where $\pi_i\left(k\right)$ denotes the payoff of an individual of type $i$ resulting from a public goods interaction with $k$ cooperators among the participating co-players.

Alternatively, the expected payoffs of individuals can be derived through ``individual centered sampling'' \citep{santos:Nature:2008}, which assumes that every individual ``hosts'' a public goods interaction with equal chances and invites $n-1$ random individuals from its neighborhood to participate. This means that each individual finds itself in three roles: \emph{(i)} hosting a public goods interaction, or participating in a public goods hosted by \emph{(ii)} a cooperator or \emph{(iii)} a defector. On average, for every interaction hosted, each individual engages in $n-1$ interactions hosted by others. In the case of well-mixed populations, the neighborhood of each individual includes everyone else in the population. Thus, the payoff contributions for a cooperator hosting an interaction, $u_C^\text{host}$, as well as participating in interactions hosted by other cooperators, $u_C^C$, or defectors, $u_C^D$, are given by
\begin{subequations}
	\begin{align}
		u_C^\text{host} &= \sum_{k=0}^{n-1} \frac{\binom{X-1}{k} \binom{N-X}{n-1-k}}{\binom{N-1}{n-1}}\pi_C\left(k+1\right) ;\\
		u_C^C &= (n-1)\frac{X-1}{N-1} \sum_{k=0}^{n-2} \frac{\binom{X-2}{k} \binom{N-X}{n-2-k}}{\binom{N-2}{n-2}}\pi_C(k+2)\notag\\
		&= \sum_{k=0}^{n-1} k\frac{\binom{X-1}{k} \binom{N-X}{n-1-k}}{\binom{N-1}{n-1}}\pi_C\left(k+1\right) ;\\
		u_C^D &= (n-1)\frac{N-X}{N-1} \sum_{k=0}^{n-2} \frac{\binom{X-1}{k} \binom{N-X-1}{n-2-k}}{\binom{N-2}{n-2}}\pi_C\left(k+1\right)\notag\\
		&= \sum_{k=0}^{n-1} (n-1-k)\frac{\binom{X-1}{k} \binom{N-X}{n-1-k}}{\binom{N-1}{n-1}}\pi_C\left(k+1\right) .
	\end{align}
\end{subequations}
Similarly, for defectors we obtain
\begin{subequations}
	\begin{align}
		u_D^\text{host} &= \sum_{k=0}^{n-1} \frac{\binom{X}{k} \binom{N-X-1}{n-1-k}}{\binom{N-1}{n-1}}\pi_D\left(k\right) ; \\
		u_D^C &= (n-1)\frac{X}{N-1} \sum_{k=0}^{n-2} \frac{\binom{X-1}{k} \binom{N-X-1}{n-2-k}}{\binom{N-2}{n-2}}\pi_D\left(k+1\right)\notag\\
		&= \sum_{k=0}^{n-1} k \frac{\binom{X}{k} \binom{N-X-1}{n-1-k}}{\binom{N-1}{n-1}}\pi_D\left(k\right) ; \\
		u_D^D &= (n-1)\frac{N-X-1}{N-1} \sum_{k=0}^{n-2} \frac{\binom{X}{k} \binom{N-X-2}{n-2-k}}{\binom{N-2}{n-2}}\pi_D\left(k\right)\notag\\
		&= \sum_{k=0}^{n-1} (n-1-k)\frac{\binom{X}{k} \binom{N-X-1}{n-1-k}}{\binom{N-1}{n-1}}\pi_D\left(k\right).
	\end{align}
\end{subequations}
The overall payoffs for cooperators, $u_C=u_C^\text{host}+u_C^C+u_C^D$, and defectors, $u_D=u_D^\text{host}+u_D^C+u_D^D$, turn out to be the same as those for interaction groups sampled uniformly at random, c.f. \eq{randomsampling}, up to a factor of $n$. Hence, no distinctions need to be made in terms of sampling procedures in well-mixed populations. However, we note that individual centered sampling immediately and naturally translates to structured populations with limited local interactions.

The distinction between the three different roles for each individual (or, more generally, $d+1$ roles with $d$ strategic types) is particular to public goods interactions in groups with $n\geq3$. In the case of pairwise interactions ($n=2$) the payoff accrued from hosting the interaction is the same as the sum of participating in interactions hosted by cooperators and defectors, $u_C^\text{host}=u_C^C+u_C^D$ and $u_D^\text{host}=u_D^C+u_D^D$, such that the overall payoff is simply twice the expected payoff from hosting an interaction, $u_C=2 u_C^\text{host}$, $u_D=2 u_D^\text{host}$.

In finite populations and in the absence of mutations, the two homogeneous states with only defectors ($X=0$) or only cooperators ($X=N$) are absorbing. This means, in the long run, any population eventually reaches one of those two states. Relevant quantities to characterize the evolutionary dynamics in this scenario are fixation probabilities, i.e., the probability that a particular strategy eventually takes over the entire population (fixes). Of particular interest are the fixation probability of a single cooperator in a defector population, $\rho_C$, and the converse probability that a single defector fixes in a cooperator population, $\rho_D$. Just as in pairwise interactions \citep{nowak:Nature:2004,traulsen:bookchapter:2009}, the fixation probabilities can be expressed in terms of the transition probabilities $T_i^\pm$ as follows: 
\begin{subequations}
	\label{eq:rhocrhod}
	\begin{align}
		\rho_C &= \frac{1}{\displaystyle 1 + \sum_{i=1}^{N-1} \prod_{m=1}^i \frac{T_m^-}{T_m^+}} ; \\
		\rho_D &= \rho_C \prod_{m=1}^{N-1} \frac{T_m^-}{T_m^+}.
	\end{align}
\end{subequations}

The concept of fixation probabilities remains meaningful for rare mutations, $\mu N^2\ll1$, in asexually reproducing populations, or, equivalently, rare mistakes (or random exploration) when adopting strategies in cultural evolution. In this limit, the mutant strategy has either disappeared or reached fixation before the next mutant appears. In particular, this limit allows to relate the fixation probabilities, $\rho_C$ and $\rho_D$, to the probability of finding the population in one or the other absorbing state, which is proportional to the time the population spends there. More specifically, detailed balance requires $p_0\rho_C=p_N\rho_D$, where $p_0$ and $p_N$ denote the probabilities that the population is in the homogeneous defector or cooperator state, respectively. For rare mutations, the probability for all other configurations is negligible, such that $p_0+p_N\approx1$ and hence $p_N\approx\rho_C/\left(\rho_C+\rho_D\right)$ (and $p_0\approx1-p_N$).

For larger mutation rates, stationary distributions can be derived from the stochastic, tridiagonal $\left(N+1\right)\times\left(N+1\right)$ transition matrix for the modified transition probabilities $Q_i^\pm$, for uniform, \eq{q+q-unif}, or temperature-based, \eq{q+q-temp}, mutations for $i\in\left\{1,\ldots,N-1\right\}$, together with $Q_0^+=Q_N^-=\mu$ or $Q_0^+=\mu f_D/\left(f_C+f_D\right)$ and $Q_N^-=\mu f_C/\left(f_C+f_D\right)$, respectively.

\subsection{Large interaction groups}
Cooperation in larger groups becomes increasingly challenging because the impact of one individual's contribution on the performance of the public good is small. This further exacerbates global challenges such mitigating climate change. In the extreme case, $n\to N$, this means that the interaction group includes everyone in the population. The condition that it pays to switch to cooperation, \eq{pck1pdk}, becomes
\begin{align}
	\label{eq:pcX1pdX}
	\left(X+1\right) r\left(X+1\right) -X r\left(X\right) > N ,
\end{align}
which is exceedingly hard to satisfy in larger populations, or requires truly large multiplication factors. Interestingly, however, even if condition \eq{pcX1pdX} is satisfied, cooperation still does not evolve because the switch to cooperation increases the payoffs of everyone in the entire population by exactly the same amount, while $\pi_C(X)<\pi_D(X)$ holds regardless of $X$. In other words, Simpson's paradox (unfortunately) no longer applies, and the average payoff of cooperators is always lower than that of defectors because all interactions involve the entire population and hence cooperators cannot offset their losses in mixed groups through interactions in groups of only cooperators (and, similarly, defectors are not disadvantaged by interactions in groups of only defectors).

Withholding cooperation in such situations becomes an act of spite, but, paradoxically, doing so remains evolutionarily advantageous. This outcome can be illustrated using the simpler linear public goods game: for $r>N$ it pays to switch to cooperation, so $\pi_C\left(X+1\right) >\pi_D\left(X\right)$. But $\pi_C\left(X\right) -\pi_D\left(X\right) =-c$ for all $X$. Considering the ratio of the transition probabilities, \eq{t+t-moran} or \eq{t+t-pair}, both yield
\begin{align}
	\frac{T_X^-}{T_X^+} = e^{wc} > 1
\end{align}
for $w, c>0$. Hence, the propensity to lower the number of cooperators always exceeds that of increasing them.

\subsection{Linear multiplication factors}
For linearly increasing multiplication factors, \eq{rkr1rn}, the expected payoffs for cooperators and defectors in \eq{randomsampling} simplify to
\begin{subequations}
	\begin{align}
		u_C &= \left(r_1-1\right) c -\left(N-X\right) B\left(X\right) -S\left(X\right)\left(X-2\right) c ; \\
		u_D &= X \left(B\left(X\right) -S\left(X\right) c\right) ,
	\end{align}
\end{subequations}
with
\begin{subequations}
	\begin{align}
		S\left(X\right) &= \frac{X-1}{\left(N-2\right)\left(N-1\right)}\left(r_1-r_n\right) ;\\
		B\left(X\right) &= \frac{c}{n}\left(r_1\frac{n-1}{N-1}+2S\left(X\right)\right) .
	\end{align}
\end{subequations}
Unfortunately, all other analytical expressions quickly become unwieldy. An exception is in the limit of weak selection.

\subsubsection{Weak selection}
In the weak selection limit, the fixation probabilities in \eq{rhocrhod} simplify to 
\begin{subequations}
	\label{eq:rhocrhodweak}
	\begin{align}
		\rho_C &= \frac{1}{N} + c \frac w2 \left(\frac{\left(N-n\right) \left(2 r_n+r_1\right)}{3 N n}-\frac{N-1}{N}\right) + O\left(w^2\right) ; \\
		\rho_D &= \frac{1}{N} + c \frac w2 \left(\frac{\left(N-n\right) \left(r_1-4 r_n\right)}{3 N n}+\frac{N-1}{N}\right) + O\left(w^2\right) .
	\end{align}
\end{subequations}
A strategy $i$ is \emph{advantageous} if it has a higher fixation probability than a neutral mutant, $\rho_i>1/N$, and \emph{favored} if $\rho_i>\rho_j$ \citep{nowak:Nature:2004}. Thus, cooperation is advantageous if
\begin{align}
	2r_n+r_1 &> \frac{3n \left(N-1\right) n}{N-n}.
\end{align}
In the limit of large populations, this condition reduces to $2r_n+r_1>3n$. For economies of scale ($r_n>r_1$), this condition is equivalent to $x^\ast<1/3$, where $x^\ast$ refers to the interior equilibrium of the deterministic dynamics (see \eq{xast}). For diminishing returns ($r_n<r_1$), the equivalent condition is $x^\ast>1/3$. Similarly, defection is advantageous if
\begin{align}
	4 r_n - r_1 &< \frac{3n \left(N-1\right)}{N-n},
\end{align}
which simplifies to $4 r_n - r_1 < 3 n$ or $x^\ast>2/3$ for economies of scale and large populations, while for diminishing returns the equivalent condition is $x^\ast<2/3$. As a consequence, the $1/3$-rule \citep{nowak:Nature:2004} naturally extends to group interactions.

Moreover, both strategies are advantageous when 
\begin{align}
	4 r_n - r_1 &< \frac{3n \left(N-1\right)}{N-n} < 2r_n+r_1.
\end{align}
This inequality implies $r_n>r_1$ and hence only applies to economies of scale. With reversed inequalities, neither strategy is advantageous, which implies $r_n<r_1$ and hence requires diminishing returns.

\begin{figure}[tp]
	\centering
	\includegraphics[width=0.6\textwidth]{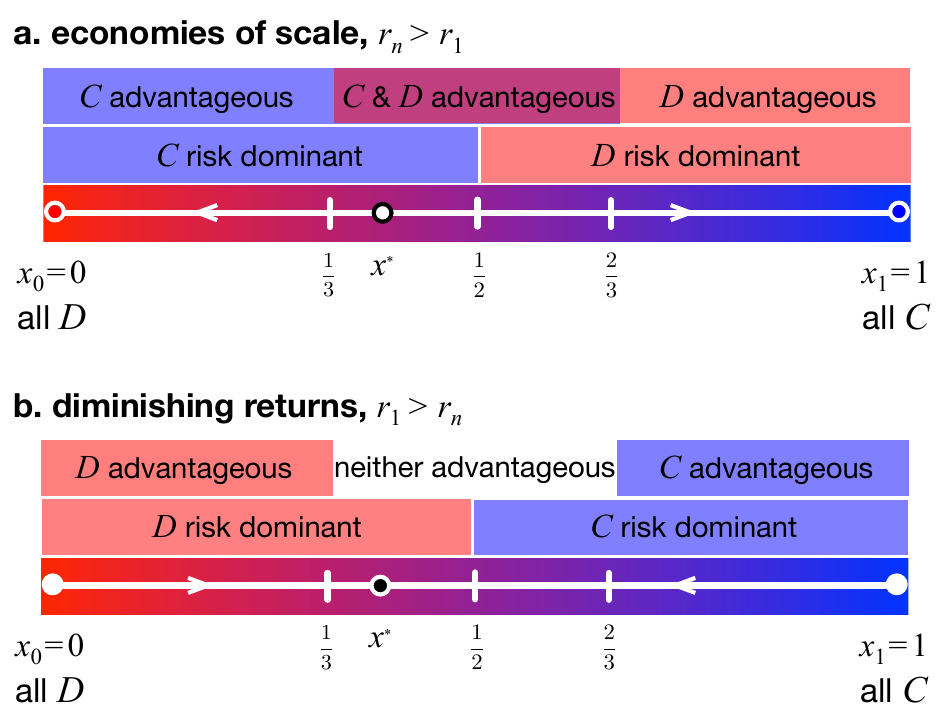}
	\caption{\label{fig:advfav}
		Summary of the relationships between the interior equilibrium, $x^\ast$, and fixation probabilities of cooperators and defectors, $\rho_C$ and $\rho_D$, in nonlinear public goods games for \bpanel{a} economies of scale ($r_n>r_1$) and \bpanel{b} diminishing returns ($r_n<r_1$). If the fixation probability of a strategy exceeds that of a neutral mutant, the strategy is \emph{advantageous} and if it exceeds that of the other type, it is \emph{favored} (risk dominant). In the limit of large populations and weak selection the location of $x^\ast$ determines which strategies are advantageous, if any, and which are favored.}
\end{figure}

Cooperation is favored ($\rho_C>\rho_D$) for sufficiently high returns of universal cooperation, $r_n>n\left(N-1\right) /\left(N-n\right)$, or simply $r_n>n$ for large populations. Note that after subtracting $r_1$ on both sides, the latter inequality is equivalent to $x^\ast<1/2$ for $r_n>r_1$ and the reverse for $r_n<r_1$. This recovers the condition for risk dominance, which denotes the strategy with the larger basin of attraction. For a graphical summary, see \fig{advfav}.

\subsubsection{Stationary distributions}
\begin{figure}[tp]
	\centering
	\includegraphics[width=0.8\textwidth]{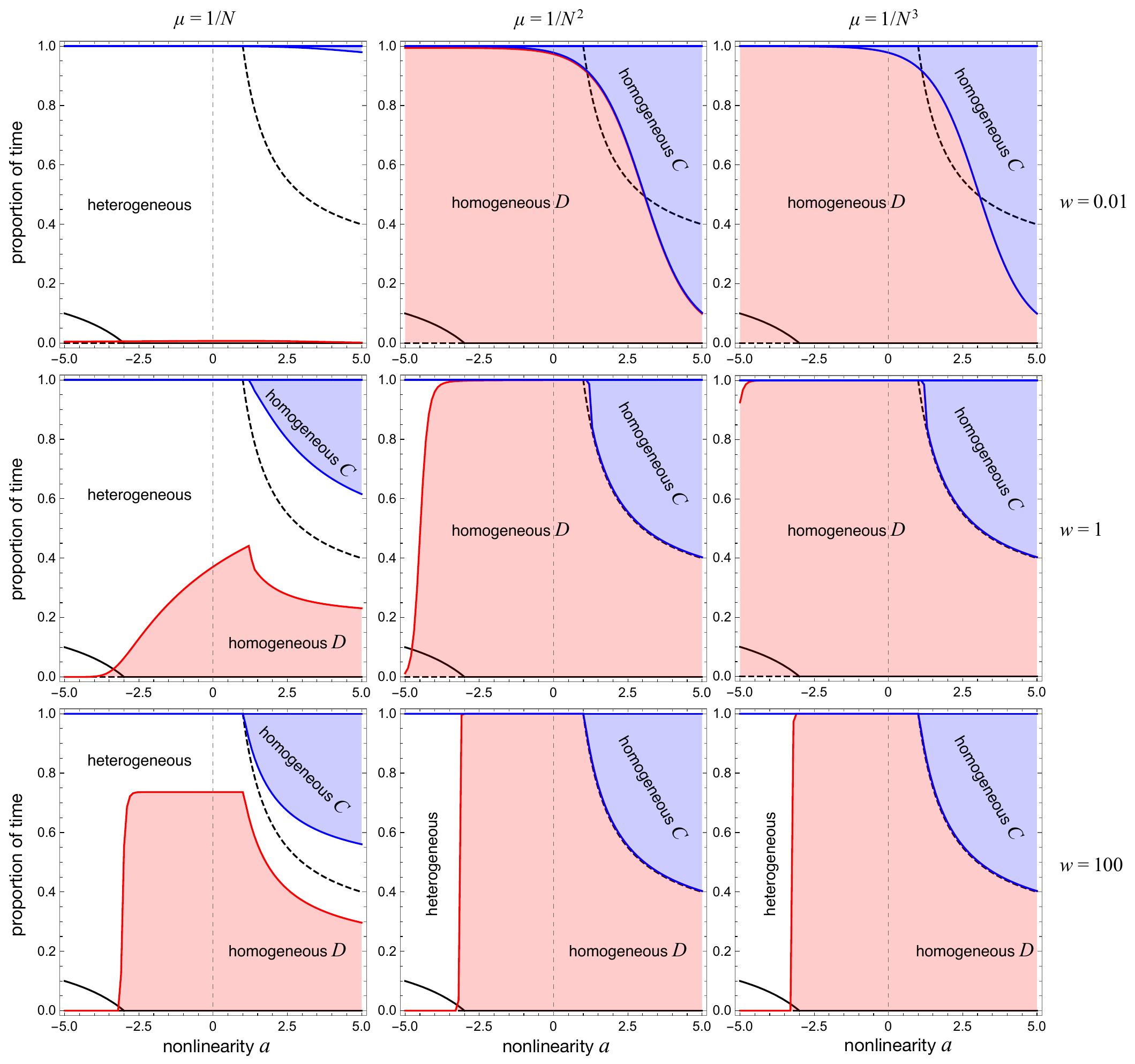}
	\caption{\label{fig:dyn:stationary}
		Relative time spent in homogeneous $C$ (blue) or $D$ (red) states (at most one individual of the opposite type) and heterogeneous mixtures of $C$ and $D$ (white) in the frequency-dependent Moran process with exponential payoff-to-fitness mapping as a function of the nonlinearity $a$ for different selection strengths (top: $w=0.01$; middle: $w=1$; bottom: $w=100$) and mutation rates (left: $\mu=1/N$; center: $\mu=1/N^2$; right: $\mu=1/N^3$). The vertical dashed line at $a=0$ separates the domain of diminishing returns (left) from economies of scale (right). Black lines indicate the equilibria (solid: stable; dashed: unstable) of the deterministic dynamics in the limit $N\to\infty$. The parameters are $n=8$, $r_1=r-a$, and $r_n=r+a$, with $r=5$ and $c=1$.}
\end{figure}

\fig{dyn:stationary} depicts the (numerically calculated) stationary distribution as a function of the nonlinearity, $a$, with $r_1=r-a$ and $r_n=r+a$ and for different selection strengths ($w\in\left\{0.01, 1, 100\right\}$), as well as mutation rates ($\mu\in\left\{1/N, 1/N^2, 1/N^3\right\}$). 
For neutral drift ($w=0$), the average fixation time of a single mutant is approximately $N^2$ updates (or $N$ generations). Thus, for mutation rates $\mu<1/N^{2}$ the population typically relaxes into one of the homogeneous states before a subsequent mutation arises. This provides a useful reference for whether or not mutations are rare. It also implies that once a mutant arises, it stays for approximately $N$ updates in the population, which heavily skews results in favor of heterogeneous states, especially for larger mutation rates. Thus, we assume a state is ``homogeneous'' if there is at most one individual of the other type.

For weak selection, the dynamics are dominated by random drift and the population composition is mostly determined by mutations (see the top row of \fig{dyn:stationary}): with frequent mutations ($\mu=1/N$), the state is almost always heterogeneous, whereas for less frequent mutations ($\mu<1/N^2$), noise consistently drives the population into an absorbing state.

The argument that mutants are rare for $\mu<1/N^{2}$ does not extend to stronger selection (see the bottom row of \fig{dyn:stationary}). Instead, in coexistence games ($a<r-n$) fixation times become astronomical, and the stationary state remains heterogeneous. In contrast, fixation can be much faster in coordination games ($a>\left(n-r\right) /3$) because essentially all invasion attempts fail. Interestingly, this is actually the root cause for numerical issues. Even though the stationary distribution should not depend on the initial configuration, numerical results are susceptible to them. More specifically, to prevent numerical issues we consistently used a uniform distribution as the initial configuration (but no probability mass on the absorbing states) for all panels in \fig{dyn:stationary}.

Strictly speaking, the strength of selection is to be understood in relation to the population size, given by $w N$, where larger $w$ enhances the impact of fitness differences and larger $N$ reduces the stochastic effects of demographic noise. The limit of strong selection, $w N\to\infty$ recovers the deterministic dynamics of \eql{rep:inf}{rep:moran}.

Repeating the same analysis for temperature-based mutations (\eq{q+q-temp}) does not result in any significant differences apart from a moderate increase in heterogeneity for strong selection and large mutations (not shown).

\section{Discussion}
The prisoner's dilemma is a leading mathematical metaphor to model the evolution of cooperation for interactions in pairs. Similarly, the public goods game (and a related formulation called the ``$n$-person prisoner's dilemma'' \citep{schelling1978micromotives,boyd:JTB:1988,hauert:PRSB:1997}) plays the same role for interactions in groups of arbitrary size, $n$. In the case of constant multiplication factors of the common pool, yielding the linear public goods game, the two are very similar. More specifically, each linear public goods interaction can be mapped onto $n-1$ donation game interactions, a special and particularly intuitive instance of the prisoner's dilemma where cooperators provide a benefit to their co-player at a cost to themselves.

The linear public goods game has attracted considerable interest in both theoretical \citep{hauert:Science:2002,santos:Nature:2008} and experimental studies \citep{fehr:Nature:2002,gaechter:Science:2008,yamagishi:JPSP:1986}. This is in stark contrast to any real life public goods interaction \citep{ostrom:CUP:1999,kraak:FF:2011}, where the return of the common good is almost certainly not linear. Instead, the rate of return typically increases (or decreases) with the number of contributors to represent economies of scale (or diminishing returns). Here, we present an in-depth discussion of more general, nonlinear public goods games. We cover decisions of the classical, rational player and, in order to analyze evolutionary dynamics, we discuss mappings of game payoffs to fecundity or fitness and derive transition probabilities for microscopic changes in a population of individuals both for asexual, clonal reproduction as well as through imitation or learning (cultural evolution). These transition probabilities may include mutational processes and provide a transparent, mechanistic derivation for the evolutionary outcomes of the stochastic dynamics in finite populations as well as the deterministic dynamics in the limit of infinite populations.

For finite populations, we derive the fixation probabilities, $\rho_C$ and $\rho_D$, of cooperators and defectors, together with the conditions that a strategy is advantageous ($\rho_i>1/N$) or favored ($\rho_i>\rho_j$) as well as the stationary distributions for scenarios that include mutations and hence without absorbing states. For rare mutations and increasing strengths of selection (or larger population sizes), the dynamics gradually recover the deterministic limit.

In contrast to the traditional linear public goods game, defection may no longer be a dominant strategy and the demise of cooperation no longer the inevitable outcome. Instead, richer types of interactions can be modeled based on how the multiplication factor of the common pool, $r\left(k\right)$, depends on the number of contributors, $k$. Two particularly important cases are diminishing returns for decreasing $r\left(k\right)$ and economies of scale for increasing $r\left(k\right)$. In the former case, cooperators and defectors may stably coexist (an analogue of the snowdrift game in pairwise interactions), while the latter may result in bistability, leading to universal cooperation or defection depending on the initial configuration (an analogue of the stag-hunt game for pairwise interactions).

In the linear public goods game, the value of the public pool, before being divided, is $v\left(k\right) =rkc$ when there are $k$ cooperators. The notions of synergy and discounting of cooperation, where the value of the pool takes the form $v\left(k\right) =rc\left(1+\delta +\delta^{2}+\cdots +\delta^{k-1}\right)$ were previously considered \citep{hauert:JTB:2006a}. More conceptually, this function represents the unique expression where the marginal values of an additional cooperator, $m\left(k\right)\coloneqq v\left(k+1\right) -v\left(k\right)$, form a geometric sequence. That is, $m\left(k\right) =\delta m\left(k-1\right)$ for some factor $\delta >0$, which represents synergy when $\delta >1$ and discounting when $\delta <1$. In the present approach, we have $v\left(k\right) =r\left(k\right) kc$, where $r\left(k\right)$ is a linear function of $k$. The marginals for this function form an arithmetic sequence, with $m\left(k\right) =\delta+m\left(k-1\right)$ and $\delta=2\frac{r_{n}-r_{1}}{n-1}$. In particular, the marginal value $m\left(k\right)$ is a linear function of $k$ whenever each new cooperator contributes to the pool via interactions with each of the $k$ existing cooperators. This interpretation is consistent with Metcalfe's law \citep{metcalfe:IEEE:2013}, which states that the ``value'' of a network (e.g., financial or telecommunications) grows in proportion to the square of the network size. Assuming cooperators form a network, this is equivalent to the marginal value growing linearly in $k$. This kind of growth is also reasonable in vaccination campaigns and collaborative research efforts.

A broader goal here is to emphasize the importance of including nonlinearities into mathematical models of public goods and highlight the qualitative changes in the dynamics that can arise when nonlinearities enter into public investment pools \citep{archetti:Games:2018}. In contrast to earlier studies in evolutionary dynamics \citep{chen:PRE:2012b} and economics \citep{allouch:JET:2015,allouch:JPET:2019} that allow for more general (nonlinear) benefit functions, our analysis here is on a particularly concrete example, in which the nonlinearity is characterized by essentially just one parameter (c.f. Fig.~\ref{fig:dyn:cnpgg}). There are certainly other formulations worthy of study in evolutionary settings, but already our simple setup demonstrates a notable departure from the dynamics observed in more common models of linear games.

\subsubsection*{Acknowledgements}
C.H. acknowledges funding by the National Science and Engineering Research Council Canada (NSERC), grant RGPIN-2021-02608.

\end{document}